\documentclass{iopart} 
\usepackage{graphicx}
\usepackage{epsfig}
\usepackage{color}

\usepackage{iopams}
\usepackage{amssymb}
\usepackage{amscd}
\usepackage{eucal}
\usepackage{bm}
\newcommand{\ignore}[1]{\relax}

\newcommand{\half}{{\textstyle{\frac{1}{2}}}}
\newcommand{\third}{{\textstyle{\frac{1}{3}}}}

\newcommand{\al}{\alpha}

\newcommand{\de}{\delta}

\newcommand{\om}{\omega}

\newcommand{\boxit}[1]{\vbox{\hrule height1pt \hbox{\vrule 
width1pt\kern3pt\vbox
{\kern3pt#1\kern3pt}\kern3pt\vrule width1pt}\hrule height1pt}}
\newcommand{\invisibleboxit}[1]{\vbox{ height1pt \hbox{ width1pt\kern3pt\vbox
{\kern3pt#1\kern3pt}\kern3pt\vrule width1pt} height1pt}}

\newcommand{\rhosys}{\hat{\rho}}
\newcommand{\Hsys}{\hat{\cal H}_{\rm sys}}
\newcommand{\Henv}{\hat{\cal H}_{\rm env}}
\newcommand{\Huniv}{\hat{\cal H}_{\rm univ}}

\newcommand{\sigxp}{\hat{\sigma}_{x'}}
\newcommand{\sigyp}{\hat{\sigma}_{y'}}
\newcommand{\sigz}{\hat{\sigma}_{z}}

\begin{document}

% syntax : \title[short title]{full title}
\title[
Staying positive: 
\dots perturbative master equations]
{Staying positive: 
going beyond Lindblad with perturbative master equations}

\author{Robert S. Whitney}
\address{Institut Laue-Langevin, 6 rue Jules Horowitz, B.P. 156,
         38042 Grenoble, France.}
\date{\today}

\begin{abstract}
The perturbative master equation (Bloch-Redfield) 
is extensively used to study dissipative quantum 
mechanics --- particularly for qubits ---
despite the 25 year old criticism that it violates positivity 
(generating negative probabilities). 
We take an arbitrary system coupled to 
an environment containing many degrees-of-freedom, 
and cast its perturbative master equation
(derived from a perturbative treatment of Nakajima-Zwanzig 
or Schoeller-Sch\"on equations) 
in the form of a Lindblad master equation.
We find that the equation's parameters are {\it time-dependent}.
This time-dependence is rarely accounted for, and
invalidates Lindblad's dynamical semigroup analysis.
We analyze one such  Bloch-Redfield master equation 
(for 
a two-level system coupled to an environment with a 
short but non-vanishing memory time), which apparently violates positivity.
We show analytically that, once the time-dependence of the parameters is accounted for, positivity is preserved.

\end{abstract}

\maketitle

%%%%%%%%%%%%%%%%%%%%%%%%%%%%%%%%%%%%%%%%%%%%%%%%%%%%%%%%%%%%%%%%
%%%%%%%%%%%%%%%%%%%%%%%%%%%%%%%%%%%%%%%%%%%%%%%%%%%%%%%%%%%%%%%%

\section{Introduction}\label{sect:intro}

No system is truly isolated from its environment,
thus all quantum systems experience some amount of 
dissipation and decoherence \cite{book:Cohen-Tannoudji,book:open-quantum}.  
To understand the properties of real quantum systems
we must understand the effect of dissipation in quantum mechanics.
This is extremely relevant to recent works on qubits
and quantum information processing  
(quantum computing and communication).
In experiments \cite{sc-qubits,spin-qubits,trapped-atom-qubits} 
the coupling to the environment is typically not as small
as would be required to build a quantum computer.
One must understand the effect of the environment 
on a qubit, if one wishes to minimize it.

Any theory for a quantum system 
which exchanges energy and information (but not particles) 
with its environment should give
a master equation (evolution equation) for the system's density-matrix
which satisfies three basic requirements;
\begin{itemize}
\item[(i)]
preserves the Hermiticity of the density-matrix,
so all probabilities are real,

\item[(ii)] preserves the trace of the density-matrix,
then the sum of probabilities over 
any complete set of orthogonal states is one,
\item[(iii)]
preserves {\it positivity}.
A system is positive only if
the probability of {\it all} possible states is positive.
Given (ii), this guarantees that all probabilities 
lie between zero and one. 
In this work we do not consider {\it complete positivity},
excepting comments in Sections~\ref{sect:Lindblad} and \ref{sect:concl}.

\end{itemize}
There are only a small number of models for which such master equations
can be derived exactly (we will not address these here). 
In all other cases, there are
two main methods for finding such a master equation \cite{book:open-quantum};
\begin{itemize}
\item
{\it Phenomenological method.} Here one attempts to construct general
master equations which satisfy requirements (i-iii).
Under the assumption that the evolution is translationally 
invariant in time 
(a dynamical semigroup property), 
as is often the case for Markovian evolution,
Lindblad \cite{Lindblad,semigroup-book} considered the master equation 
given in Eqs.~(\ref{eq:Lindbladeqn-a},\ref{eq:Lindbladeqn-b}).
He proved that it is the {\it most general} equation
that satisfies (i-ii) above, while also preserving complete positivity.
{\it Complete positivity} is as strong or stronger than {\it positivity},
thus it automatically satisfies (iii) above 
(see the comment in Section~\ref{sect:concl} due to \cite{Hall08}).

\item
{\it Perturbative method} \cite{book:Cohen-Tannoudji}. 
Here one takes the evolution of a 
system and its environment (from their combined Hamiltonian), 
and traces over the environment degrees-of-freedom.  
Various methods of doing this exist;
Bloch-Redfield\cite{Bloch57,Redfield57}, 
Nakajima-Zwanzig\cite{Nakajima58,Zwanzig60}, 
Schoeller-Sch\"on\cite{Schoeller94}.
However one is typically forced
to treats the system-environment interaction 
perturbatively,
then all these approaches reduce to Bloch-Redfield's.
\end{itemize}
The Lindblad master equation (the most general generator of
a dynamical semigroup) takes the form;
\numparts
\begin{eqnarray}
\fl {\rmd \over \rmd t} \rhosys (t)
&=& -\rmi \big[ {\cal H}_{\rm sys},\rhosys (t) \big]_-
- \sum_{n=1}^{N^2-1} {\lambda_n \over 2}
\Big(\hat{L}_n^\dagger\hat{L}_n\rhosys(t) +\rhosys(t)\hat{L}_n^\dagger \hat{L}_n
-2 \hat{L}_n\rhosys(t) \hat{L}_n^\dagger\Big),
\label{eq:Lindbladeqn-a}\\
\fl & & \qquad\qquad\qquad\qquad\qquad\qquad\qquad
\hbox{ with } \lambda_n \geq 0  \hbox{ for all } n, 
\label{eq:Lindbladeqn-b}
\end{eqnarray}
\endnumparts
where the commutator $[\hat A,\hat B]_-= \hat A \hat B- \hat B \hat A$,
and $\{\hat{L}_n\}$ is a set of ortho-normal ({\it trace-class}) operators.
It is often assumed that all {\it Markovian} master equations
fall into the category of dynamical semigroup evolution,
and thus Eqs.~(\ref{eq:Lindbladeqn-a},\ref{eq:Lindbladeqn-b})
give the most general Markovian evolution.
However this is a subtle point, we discuss it 
(and define terms like ``dynamical semigroup''
and ``trace-class'') in Section~\ref{sect:Lindblad}.

The perturbative method's advantage over
the phenomenological method is that 
one can study how a particular environment 
(with a given spectrum, temperature, etc) affects the system.
Thus one can address a crucial aspect of
qubit research; how should one engineer 
a particular system to minimize decoherence?
However the resulting Bloch-Redfield master equation
has long been criticized \cite{Spohn79,Spohn80},
because it can be written in the form in Eq.~(\ref{eq:Lindbladeqn-a})
but then typically violates Eq.~(\ref{eq:Lindbladeqn-b}).
In these cases it violates Lindblad's condition for complete positivity.
Further, there is plenty of evidence 
that it also violates positivity (see Section \ref{sect:lit-review}).

\subsection{Outline of this article}

The objective of this article is to study 
this apparent contradiction between the perturbative method
and Lindblad's proof.
We start by discussing, in Section~\ref{sect:Lindblad}, 
the assumptions that underlie the Lindblad master equation.
In Section~\ref{sect:Bloch}
we consider the Bloch-Redfield equation for an arbitrary
system, and show that, in general, one coupling constant, $\lambda_2$, 
is negative. 
However we also show that the parameters of the 
Bloch-Redfield master equation, $\{\lambda_n\}$ and $\{\hat{L}_n\}$,
are {\it time-dependent}.
This means the master equation does not generate a dynamical semigroup.
Thus Lindblad's proof is inapplicable to the
Bloch-Redfield equation, and {\it a priori} we do not know 
whether a negative $\lambda_2$
will lead to a violation of positivity or not.

\begin{figure}
\hskip 0.1 \textwidth 
\includegraphics[width=0.8\textwidth]{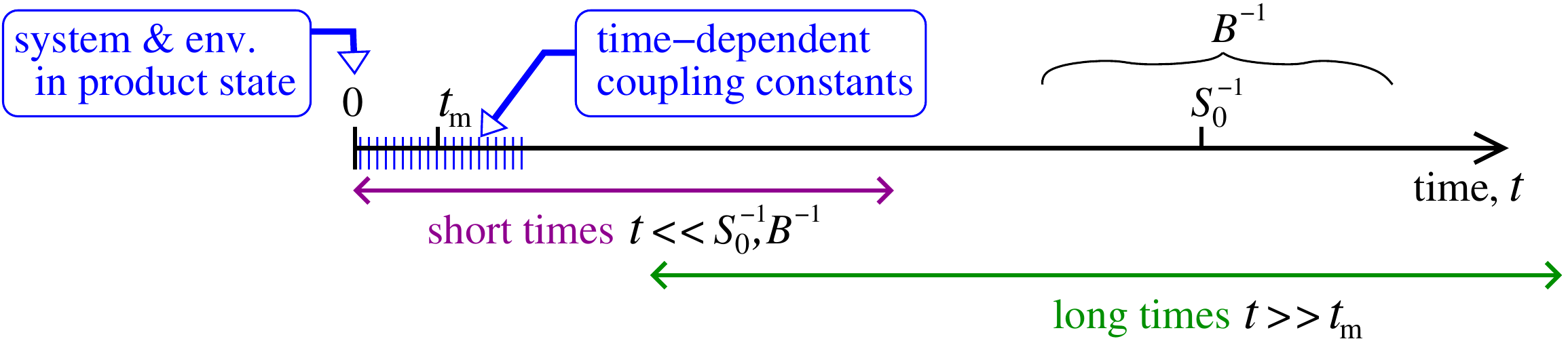}
\caption{
Timescales for the model (in Section \ref{sect:2level})
for which we show positivity, despite it not satisfying 
Lindblad's requirement, Eq.~(\ref{eq:Lindbladeqn-b}).
The system is a spin-half with Hamiltonian $-\half B\hat\sigma_z$, 
and the environment couples to it via $\hat\sigma_x$.
The environment's noise spectrum (with noise power $S_0$)
is broad, leading to a short memory time, $t_{\rm m}$. 
Decoherence and relaxation times 
(both $\sim S_0^{-1}$)
can be smaller or larger than the Larmor precession period, $B^{-1}$.
} \label{Fig:timescales}
\end{figure}

In Section~\ref{sect:2level},
we consider the Bloch-Redfield equation for a particular 
system (a two-level system coupled to an environment with a 
very broad spectrum of excitations).
We divide the evolution into two overlapping regimes; short- and long-times 
(sketched in Fig.~\ref{Fig:timescales}).
The time-dependence of the parameters is only relevant in the 
short-time regime ($t$ much less than decoherence/relaxation times).
We show analytically that the system remains positive 
in both regimes (i.e.~for all $t\geq 0$), 
despite the negative coupling constant, $\lambda_2$.

%%%%%%%%%%%%%%%%%%%%%%%%%%%%%%%%%%%%%%%%%%%%
%%%%%% figure goes here %%%%%%
%%%%%%%%%%%%%%%%%%%%%%%%%%%%%%%%%%%%%%%%%%%%
%%%%%%%%%%%%%%%%%%%%%%%%%%%%%%%%%%%%%%%%%%%%

%-------------------------------------
\subsection{The place of this work in the literature}
\label{sect:lit-review}

In traditional derivations of the 
Bloch-Redfield master equation\cite{book:Cohen-Tannoudji,book:open-quantum},
it is assumed that the  parameters of the master equation
are time-independent.
In reality all environment-induced terms in the master equation
are zero at $t=0$ (defined as the time at which the system and environment 
are in a factorized state), before growing with $t$ and saturating at 
$t \gg t_{\rm m}$, where $t_{\rm m}$ is the environment memory time.
So the assumption of time-independence is {\it flawed} 
for times of order the memory time, $t_{\rm m}$.
This has been discussed in the context of 
coupled classical oscillators \cite{Geigenmuller83}, 
over-damped Brownian motion 
\cite{Haake-Lewenstein83},
damped quantum oscillators
\cite{Haake-Reibold85,Gnutzmann-Haake96,Yu00,Maniscalco03,
footnote:non-Lindblad-form},
dissipative two-level systems 
\cite{Suarez92,Gaspard-Nagaoka99,Yu00,Cheng05}
and more generally \cite{Gorini89,Wonderen00}.
Nearly all these works consider dynamics on times of order 
the $t_{\rm m}$ as an {\it initial-slip}, after which the dynamics
is given by the time-independent master equation, the justification for
this is sketched in \ref{append:initial-slip}.
Of most relevance to us are those works 
which try to show that positivity is preserved in this context 
\cite{Suarez92,Gaspard-Nagaoka99,Yu00,Maniscalco03,Cheng05,Wonderen00}.
However these works provide only plausibility arguments 
\cite{footnote:earlier-arguments-for positivity}, or numerical studies
(they evolved a finite number of 
initial conditions and checked that negative probabilities 
did not emerge).
In contrast, for our model, we consider all {\it possible} initial conditions
and thereby prove analytically that positivity is preserved.
 
It has been noted that
course-graining can ensure positivity \cite{Lidar01}.
The work presented here 
indicates that the usual assumption of time-{\it independent} parameters
in the master equation only leads to a violation of positivity for
$t \lesssim t_{\rm m}$.
Thus course-graining on such a scale could 
hide such a violation.
It is also common to simplify Bloch-Redfield equations by making
a rotating-wave approximation \cite{Spohn79,Spohn80,Munro96}
which is also a form of course-graining since it
``averages out'' fast oscillations.
However if we treat the time-dependence of the parameters correctly,
the Bloch-Redfield equation is derived without any approximations 
which fail on short timescales, so
it should preserve positivity without any course-graining.

There has been a lot of interest in 
a particular class of non-Markovian master equations 
which are positive by construction. 
They are
either constructed by averaging Markovian master equations
\cite{non-Markovian1},
or by measurement processes \cite{Shabani-Lidar05}.
However, while these models are extremely interesting,
we are not aware of works relating them 
to microscopic models of a typical
qubit experiencing dissipation \cite{Budini-Schomerus05}.

Finally, we mention that some works suggested that
the reason for negative probabilities was the choice of factorized initial
conditions \cite{Munro96,vanKampen}.  
They argued that this initial condition was {\it unphysical},
and a more physical initial condition would not generate negative 
probabilities.  However, factorized initial conditions
correspond to any situation in which one makes a 
projective measurement of the system state at the start of the evolution.  
Thus, while other initial conditions
are worthy of study \cite{Pechukas94+comments,Shaji05} 
(and highly relevant to certain experimental protocols),
a factorized initial condition is {\it not} unphysical, and thus
should not be able to generate negative probabilities.
In this work we restrict ourselves to factorized initial conditions
(see Section \ref{sect:Bloch}).

%-----------------------------------------------------------------------

\section{The Lindblad master equation}
\label{sect:Lindblad}

The Lindblad master equation, 
Eqs.~(\ref{eq:Lindbladeqn-a},\ref{eq:Lindbladeqn-b}),  
is written in terms of 
a set of $N^2$ {\it trace-class} operators, $\{\hat{L}_n\}$
(where $N$ is the number of levels of the system).
Operators are trace-class if they 
form a complete orthonormal basis in the space of
system operators, 
with the scalar-product defined as
$(\hat{L}_i^\dagger \cdot \hat{L}_j)
\equiv \tr [\hat{L}_i^\dagger\hat{L}_j] $, see \cite{footnote:example-Ls}.
The basis is complete if any system operator can be written as
$\hat{O}_{\rm sys} = \sum_j \hat{L}_j \tr \big[\hat{L}_j^\dagger 
\hat{O}_{\rm sys}\big] $.
We choose $L_0$ to be proportional to the unit matrix.
One can see that Eq.~(\ref{eq:Lindbladeqn-a}) 
preserves the Hermiticity and trace of
the system's density-matrix
(the latter requires cyclic permutations inside the trace).
The combination of  Eq.~(\ref{eq:Lindbladeqn-b}) with  
Eq.~(\ref{eq:Lindbladeqn-a}) guarantees positivity. 
In fact it guarantees a stronger condition called 
{\it complete positivity}, which is the requirement that all probabilities
remain positive even if the system became entangled with
a second system at $t<0$, but then does not interact with it again.
For a review see Sections 2.4 and 3 
of Ref.~\cite{book:open-quantum}, section VB of Ref.~\cite{Spohn80}
or the introduction of Ref.~\cite{Shaji05}.
In this article we concern ourselves with studying positivity
not complete positivity, however it has recently been shown \cite{Hall08} 
that the two
are equivalent for the model that we study in section~\ref{sect:2level}.

Lindblad proved that Eqs.~(\ref{eq:Lindbladeqn-a},\ref{eq:Lindbladeqn-b})
give the most general dynamical semigroup evolution 
\cite{Lindblad,semigroup-book}.
However to understand if this is applicable to a given system, one 
must ask if that system has the properties of a {\it dynamical semigroup}.
For this one looks at
the density-matrix propagator ${\mathbb K}(t;t_0)$, 
which acts
on the density-matrix
at $t_0$ to give the density-matrix at time $t$, 
so in terms of matrix elements
\begin{eqnarray}
\hat{\rho}_{i'j'}(t)
=\sum_{ij}{\mathbb K}_{i'j';ij}(t;t_0) \hat{\rho}_{ij}(t_0).
\label{eq:K}
\end{eqnarray}
This super-operator, ${\mathbb K}(t;t_0)$, 
is an $N\times N\times N\times N$ tensor 
which acts on the $N\times N$ density-matrix.
Substituting it into Eq.~(\ref{eq:Lindbladeqn-a}) gives a 
master equation for ${\mathbb K}_{i'j';ij}(t)$.
The requirements for ${\mathbb K}(t;t_0)$ 
to form a dynamical semigroup are given in 
Refs.~\cite{book:open-quantum,semigroup-book,Spohn80},
they include (i-ii) above and complete positivity.
However another crucial requirement
is that the propagator must be translationally invariant in time, so
${\mathbb K}(t;t_0)={\mathbb K}(t-t_0)$ for all $t,t_0>0$ 
(where the system and environment were in a factorized state at time $t=0$).
Only then does 
$
{\mathbb K}_{i''j'';ij}(t_2+t_1) 
= \sum_{i'j'}{\mathbb K}_{i''j'';i'j'}(t_2){\mathbb K}_{i'j':ij}(t_1)$.
Thus 
a master equation must have {\it time-independent} parameters
to have this semigroup property.
If either the system Hamiltonian or the environment couplings 
(coupling constants $\lambda_n$ or operators $\hat{L}_n$) are time-dependent,
then ${\mathbb K}(t;t_0)$ is {\it not} translationally invariant in time.
Thus Lindblad's proof is inapplicable for such systems, even if their
evolution is {\it Markovian} (in the sense that
$\rmd \rhosys(t)/\rmd t$ is a function only of $\rhosys(t)$
not $\rhosys(t'<t)$).
So if $\lambda_n$ or $\hat{L}_n$ are time-dependent
(as in our perturbative analysis)
one cannot {\it a priori} state that negative $\lambda_n$
will lead to a violation of positivity.

%-----------------------------------------------------------------------

\section{The perturbative (Bloch-Redfield) master equation}
\label{sect:Bloch}

We assume that the system and environment start (at $t=0$)
in a factorized state $\rhosys(t=0) \otimes \hat \rho_{\rm env}$. 
This would be the case if the experiment started with a perfect 
projective measurement of the state of the system 
\cite{footnote:factorized-state}.
The ``universe'' (system $+$ environment) then evolves under the
Hamiltonian,
\begin{eqnarray}
\hat{\cal H}_{\rm univ} = \hat{\cal H}_{\rm sys} + \hat{\cal H}_{\rm env} 
+ \hat{\Gamma}\hat{x}, 
\label{eq:Huniv}
\end{eqnarray}
where $\hat{\Gamma}$ and $\hat{x}$ are 
system and environment operators, respectively.
We treat these operators as Hermitian, because we assume they are observables
(i.e.~charge, magnetic dipoles, etc) as is the case in most 
qubit experiments (and more generally).
Without lose of generality we can assume $\hat{\Gamma}$ is dimensionless
and $\hat{x}$ has units of energy.

For a suitable environment one can derive the Bloch-Redfield 
master equation for the evolution of the system's reduced density-matrix, 
$\rhosys(t)$, from the evolution of the universe's state (tracing out the
environment at time $t$).  
The assumptions necessary to derive this master equation
are discussed in \ref{append:Dyson->Bloch}.
Broadly speaking one needs an environment with a broad 
(almost) continuous spectrum of excitations, 
then the memory kernel of the environment 
(defined in Eq.~(\ref{eq:alpha}) below) 
decays on a timescale $t_{\rm m}$.
Typically the Bloch-Redfield master equation is valid when
the memory time,  $t_{\rm m}$, is much less than timescales
associated with dissipation (relaxation and decoherence),
which go like $1/(|\hat{\Gamma}\hat{x}|^2t_{\rm m})$
(we set $\hbar=1$ throughout this article).
The Bloch-Redfield master equation can be written as 
\numparts
\begin{eqnarray}
\label{eq:master-Bloch}
\fl
{\rmd \over \rmd t} \rhosys (t)
= -\rmi \big[ {\cal H}_{\rm sys},\rhosys (t)\big]_-
- \hat{\Gamma} \hat{\Xi} \rhosys(t) 
- \rhosys(t) \hat{\Xi}^\dagger \hat{\Gamma} 
+ \hat{\Xi} \rhosys(t) \hat{\Gamma} 
+ \hat{\Gamma} \rhosys(t) \hat{\Xi}^\dagger, 
\end{eqnarray}
with $\Gamma$ being the operator in Eq.~(\ref{eq:Huniv})
and
\begin{eqnarray}
\hat{\Xi} = 
\int_0^t \rmd \tau \al(\tau) \,
\exp [-\rmi \hat{\cal H}_{\rm sys}\tau] 
\hat{\Gamma} \exp [\rmi \hat{\cal H}_{\rm sys}\tau].
\label{eq:def-Xi}
\end{eqnarray}
Unlike many derivations we do not assume that we can take the
upper-bound on this integral to $\infty$.
The function $\al (\tau)$ is the environment's memory kernel,
given by 
\begin{eqnarray}
\al(\tau) 
&=& {\rm tr}_{\rm env} \big[
\hat{x} \,\exp [-\rmi \hat{\cal H}_{\rm env}\tau]\, \hat{x} 
\, \exp [\rmi \hat{\cal H}_{\rm env}\tau] \ \hat{\rho}_{\rm env}(t) \big].
\label{eq:alpha}
\end{eqnarray}
\endnumparts
Since $\al(\tau)$ is typically complex, 
$\hat{\Xi}$ is not usually Hermitian (unlike $\hat{\Gamma}$).
We assume 
that $\al(\tau)$ is independent of $t$, then $\al(-\tau)=\al^*(\tau)$.
This is true if the environment is large enough that
it is unaffected by the system-environment coupling
(during the experiment), {\it and} the initial 
environment state obeys  
$[{\cal H}_{\rm env},\hat{\rho}_{\rm env}]=0$.
The latter is the case if 
the environment is in an eigenstate
or a classical mixture of eigenstates 
(such as a thermal state).  
We assume that $\al(\tau)$ is a decaying function of $\tau$,
and define the memory time, $t_{\rm m}$, as the timescale of that decay.
Then $\hat{\Xi}$ is $t$-dependent, because $t$ appears
in the upper-bound on the integral in Eq.~(\ref{eq:def-Xi}). 

Eq.~(\ref{eq:master-Bloch}) looks 
Markovian, in the sense that the rate of change of $\rhosys(t)$ 
depends only on the value of $\rhosys(t)$ 
(not the 
value of $\rhosys(t')$ for $t'<t$).
Despite this memory effects are present in the memory kernel,
$\al(\tau)$.
As we see in \ref{append:Dyson->Bloch},
if $\al(\tau)$ is finite for a given $\tau$ 
it means the rate of change of $\hat{\rho}$ at time $t$ is 
affected by $\hat{\rho}(t-\tau)$.
This is the reason for the time-dependence of
$\hat{\Xi}$, which is zero at $t=0$, and grows to saturate 
on a timescale of order the environment memory time, $t_{\rm m}$.

By writing
$
\hat{\Gamma} \hat{\Xi}\rhosys -\rhosys\hat{\Xi}^\dagger \hat{\Gamma} 
=\half 
\big[(\hat{\Gamma}\hat{\Xi}+\hat{\Xi}^\dagger\hat{\Gamma}),\rhosys\big]_+
-\rmi \big[ {\textstyle {\rmi \over 2}}(\hat{\Gamma}\hat{\Xi}-\hat{\Xi}^\dagger\hat{\Gamma}),\rhosys\big]_-$,
where$[A,B]_\pm =AB\pm BA$ are the anti-commutator/commutator,
Eq.~(\ref{eq:master-Bloch}) becomes
\begin{eqnarray}
\fl {\rmd \over \rmd t} \rhosys (t)
= -\rmi \Big[\Hsys',\rhosys (t) \Big]_-
 - \half \Big[ (\hat{\Gamma}\hat{\Xi}+\hat{\Xi}^\dagger\hat{\Gamma}), 
\rhosys(t) \Big]_+
+ \hat{\Xi} \rhosys(t) \hat{\Gamma} 
+ \hat{\Gamma} \rhosys(t) \hat{\Xi}^\dagger, 
\label{eq:Bloch3}
\end{eqnarray}
where we define 
$\Hsys' \equiv \Hsys 
-\half \rmi (\hat{\Gamma}\hat{\Xi}-\hat{\Xi}^\dagger\hat{\Gamma})$.
Even when $\hat{\Xi}\neq \hat{\Xi}^\dagger$,
both $(\hat{\Gamma}\hat{\Xi}+\hat{\Xi}^\dagger\hat{\Gamma})$
and $\rmi (\hat{\Gamma}\hat{\Xi}-\hat{\Xi}^\dagger\hat{\Gamma})$ are Hermitian.
The fact that $\Hsys'$ is Hermitian means that
we can interprete it as a renormalized system Hamiltonian.

It is very convenient to define the symmetrized and anti-symmetrized
spectral function of the noise, $S(\om)$ and $A(\om)$
such that
\numparts
\begin{eqnarray}
\half [\al(\tau)+\al(-\tau)] = {\rm Re} [\al(\tau)] 
&=& \int {\rmd \om \over 2\pi} S(\om) \exp[-\rmi \om \tau], 
\label{eq:S-omega}
\\
\half [\al(\tau)-\al(-\tau)] = \rmi \, {\rm Im} [\al(\tau)] 
&=& \int {\rmd \om \over 2\pi}  A(\om) \exp[-\rmi \om \tau], 
\label{eq:A-omega}
\end{eqnarray}
\endnumparts
remembering that we set $\hbar=1$ throughout.
One can extract the form of $S(\om)$ and  $A(\om)$
from environment details (a bath of harmonic oscillators
\cite{Caldeira83+Leggett87}, 
a bath of spins \cite{Prokofev-Stamp}, etc).
For an environment in thermal equilibrium at temperature $T$
\cite{Makhlin-review03}, 
$S(\om)$ and $A(\om)$ are related via
$A(\om) = S(\om)\tanh (\om /2k_{\rm B} T)$ 
\cite{footnote:relation-S-to-A}.
For harmonic oscillators, $A(\om)\propto J(\om)$ and so
$S(\om) \propto J(\om) \coth (\om /2k_{\rm B} T)$, 
where $J(\om)$ is the  spectral-density in Ref.~\cite{Caldeira83+Leggett87}.

\subsection{Dephasing and Lamb shift when 
a rotating-wave approximation is reasonable}

When the dynamics is dominated by the system Hamiltonian
(off-diagonal matrix elements decay over many Larmor oscillations),
then we can make a rotating-wave (or secular) approximation 
\cite{Landau-Lif-chapter} of Eq.~(\ref{eq:Bloch3}).
We write $\rhosys(t)$ in the eigenbasis ${\cal H}_{\rm sys}$ (so  
${\cal H}_{{\rm sys};ij} = E_i \de_{ij}$),
then we can expect  
$\hat\rho^{\rm rot}_{ij}(t)=\e^{\rmi (E_i-E_j)t} \hat\rho_{{\rm sys};ij}(t)$
to be insensitive to all fast oscillating contributions
to its dynamics.
We neglect (``average out'')
contributions to $(\rmd \hat\rho^{\rm rot}_{ij}/\rmd t)$
which come from $\hat\rho_{{\rm sys};i'j'}$ when $i'\neq i$ or $j' \neq j$,
since these contributions oscillate fast, at a rate $(E_{i'}-E_{j'}-E_i+E_j)$
\cite{footnote:maxime}.
Then $(\rmd/\rmd t) \hat\rho^{\rm rot}_{ij} 
= \big[\rmi {\Delta E}(i,j)-T_2^{-1}(i,j)\big]
\hat\rho^{\rm rot}_{ij}$.
The dephasing rate, at which a super-position of states $i$ and $j$
decays to a classical mixture
($1/T_2$ for two-level systems)
is 
\begin{eqnarray}
\fl
T_2^{-1}(i,j)
&\simeq& 
{\rm Re} \Big[
\half(\hat{\Gamma}\hat{\Xi}+\hat{\Xi}^\dagger\hat{\Gamma})_{ii}
\, + \, 
\half (\hat{\Gamma}\hat{\Xi}+\hat{\Xi}^\dagger\hat{\Gamma})_{jj} 
\, -\,  
\hat{\Xi}_{ii} \hat{\Gamma}_{jj} 
\, -\, 
\hat{\Gamma}_{ii} \hat{\Xi}^\dagger_{jj} \Big].
\label{eq:T_2}
\end{eqnarray}
The coupling to the environment also causes a Lamb shift;
the precession rate is modified by the sum of the
modification in $\Hsys'$ and ${\Delta E}(i,j)$, 
where ${\Delta E}(i,j)$ is the imaginary part of the
square brackets in Eq.~(\ref{eq:T_2}).

%-------------

\subsection{Writing the Bloch-Redfield equation as a Lindblad equation}
\label{sect:Bloch-2-Lindblad}

To cast Eq.~(\ref{eq:Bloch3}) 
in the Lindblad form, we rewrite it 
in terms of a set of orthonormal (trace-class) 
operators, $\{\hat{P}_i\}$.
We use the usual Gram-Schmidt procedure;
defining $\hat{P}_1 \propto \hat{\Gamma}$,
and $\hat{P}_2$ as proportional to the component of
$\hat{\Xi}$ which is orthogonal to $\hat{\Gamma}$.
The constants of proportionality are such that both 
 $\hat{P}_1$ and  $\hat{P}_2$ are normalized. 
Hence 
\begin{eqnarray}
\hat{P}_1 \, =\, 
{\hat{\Gamma} \over \sqrt{\tr [\hat{\Gamma}^2]}}, 
\nonumber \\
\hat{P}_2 
\, = \,
{\hat{\Xi} -   \hat{P}_1\,\tr [\hat{P_1}^\dagger\hat{\Xi}]
\over
\sqrt{\tr[\hat{\Xi}^\dagger\hat{\Xi}] - 
\big|\tr [\hat{P_1}^\dagger\hat{\Xi}] \big|^2 }},
\label{eq:P_12}
\end{eqnarray}
so $\hat{P}_1$ is Hermitian while in general $\hat{P}_2$ is not.
As $\hat{P_1},\hat{P_2}$ form an orthonormal
basis, we have 
$\hat \Gamma 
=\hat{P_1} \tr [\hat{P_1}\hat\Gamma] + \hat{P_2} \tr [\hat{P_2}\hat\Gamma]$
and   
$\hat \Xi 
=\hat{P_1} \tr [\hat{P_1}\hat \Xi] + \hat{P_2} \tr [\hat{P_2}\hat \Xi]$.
Then the Bloch-Redfield equation becomes
\begin{eqnarray}
\label{eq:Bloch-nearlyLindblad}
\fl
{\rmd \over \rmd t} \rhosys (t)
&=& -\rmi[{\cal H}'_{\rm sys},\rhosys (t)]
\, - \,
\half \sum_{ij}h_{ij} 
\big(
\hat{P}^\dagger_i\hat{P}_j \rhosys(t) 
\, +\,
\rhosys(t) \hat{P}^\dagger_i\hat{P}_j 
\, -\, 2  
\hat{P}_j \rhosys(t) \hat{P}^\dagger_i  \big).
\end{eqnarray}
In general, 
$h_{ij} = \tr [\hat\Gamma^\dagger \hat{P_i}] \tr [\hat{P_j}^\dagger\hat{\Xi}]
+\tr [\hat \Xi^\dagger \hat{P_i}] \tr [\hat{P_j}^\dagger\hat{\Gamma}]$. 
However here $\tr [\hat{P}_2^\dagger\hat{\Gamma}]=0$, 
so $h_{ij}$ is given by the $ij$th element of the matrix
\begin{eqnarray}
{\bf h} 
&=& b_z  
\left( \begin{array}{cc} 
1&0 \\ 0&1
\end{array}\right)
+\left( \begin{array}{cc} 
b_z& b_+ \\ b_+^* &-b_z
\end{array}\right)
\label{eq:def-b_i},
\end{eqnarray}
where for the compactness of what follows we have defined
\begin{eqnarray}
b_+ = 
\tr [\hat{\Gamma}^\dagger \hat{P_1}]\tr [\hat{P_2}^\dagger \hat{\Xi}],
\nonumber \\
b_z ={\rm Re}\big( \tr [ \hat{\Gamma}^\dagger \hat{P}_1]
\tr [\hat{P}_1^\dagger \hat{\Xi}] \big), 
\label{eq:bs}
\end{eqnarray}
we also define $b^2= |b_+|^2+b_z^2$. 
We retain  $\dagger$s on the symbols to make the structure clear,
however $\hat{\Gamma}^\dagger=\hat{\Gamma}$
and  $\hat{P}_1^\dagger=\hat{P}_1 \propto \hat{\Gamma}$.
The eigenvalues, $\lambda_{1,2}$, and the SU(2) rotation, ${\cal U}$, 
to the eigenbasis of ${\bf h}$, are
\begin{eqnarray}
\lambda_{1,2}
&=& b_z \pm b, 
\nonumber \\ 
{\cal U} &=& 
{1\over\sqrt{2}}
\left( \begin{array}{cc}
(1 +b_z/b)^{1/2} & 
{ b_+\over \sqrt{b(b +b_z)} }\\ 
{ b_+^*\over \sqrt{b(b +b_z)} }&
-(1 +b_z/b)^{1/2}
\end{array} \right).
\label{eq:lambda-and-U}
\end{eqnarray}
Performing this rotation on Eq.~(\ref{eq:Bloch-nearlyLindblad}),
the Bloch-Redfield equation takes the form of the Lindblad equation, Eq.~(\ref{eq:Lindbladeqn-a}),
with $\hat{L}_i = \sum_{j=1,2}{\cal U}_{ij}\hat{P}_j$ 
\cite{footnote:Hsys-primed}. 
However in general $\lambda_2$ is negative \cite{Spohn79,Spohn80},
so this does {\it not} satisfy
Lindblad's requirement in Eq.~(\ref{eq:Lindbladeqn-b}).

%-------------------------------------------------------------------------
\section{Perturbative master equation for an extremely short memory time}
\label{sect:short-mem}

We assume here that the memory time, $t_{\rm m}$, 
is much shorter than any timescale in ${\cal H}_{\rm sys}$,
i.e.~$t_{\rm m} \ll \Delta_{\rm sys}^{-1}$ 
where $\Delta_{\rm sys}$ is the largest energy difference
in the system's spectrum.
We substitute 
$\hat{\Gamma}(-\tau) = 
\hat{\Gamma} -\rmi [\hat{\cal H}_{\rm sys},\hat{\Gamma}]_-\tau -\half 
[\hat{\cal H}_{\rm sys},[\hat{\cal H}_{\rm sys},\hat{\Gamma}]_- ]_- \tau^2
+ {\cal O}\big[(\Delta_{\rm sys}\tau)^3\big]$
into $\hat{\Xi}$.
We expect that $\al(\tau)$ is always given by
a dimensionless function of $\tau/t_{\rm m}$ multiplied by $t_{\rm m}^{-2}$ 
(given that $\hbar=1$).  Then
$\hat{\Xi}(t)$ (having units of energy) is 
\begin{eqnarray}
\fl
\hat{\Xi}(t)&=& f_0(t)\, \hat{\Gamma} 
\, -\, \rmi t_{\rm m} f_1(t) \,[\hat{\cal H}_{\rm sys},\hat{\Gamma}]_- 
\nonumber \\
\fl
& & \, -\, \half t_{\rm m}^2 f_2(t) 
\,[\hat{\cal H}_{\rm sys},[\hat{\cal H}_{\rm sys},\hat{\Gamma}]_-]_- 
\ +\  {\cal O}\big[t_{\rm m}^{-1}(\Delta_{\rm sys}t_{\rm m})^3\big],
\label{eq:Xi-short}
\end{eqnarray}
where
$f_q(t)= \int_0^t \rmd \tau \  (\tau/t_{\rm m})^q \ \al(\tau)$.
For all $q$, $f_q(t)$ goes like $t_{\rm m}^{-1}$ multiplied 
by a dimensionless function of $t/t_{\rm m}$.
Thus Eq.~(\ref{eq:Xi-short}) is an expansion to second-order in powers of 
$\Delta_{\rm sys}t_{\rm m}$.
Writing $f_q(t)$ in terms of $S(\om)$ and $A(\om)$ we have
\begin{eqnarray}
f_q(t) = {\rmi \over (\rmi t_{\rm m})^q} \int {\rmd \om \over 2 \pi} 
\big(S(\om)+A(\om) \big) {\rmd^q \over \rmd \om^q} 
\left[{ 1-\e^{\rmi \om t} \over \om+\rmi 0^+}\right],
\label{eq:f_q-S-and-A}
\end{eqnarray}
where a positive infinitesimal constant, $0^+$, 
ensures the convergence for $t\to \infty$. 
Thus
\begin{eqnarray}
\fl
\hat{\Xi}(t) 
=
\Big(f_0(t) \sqrt{ \tr [\hat\Gamma^2]} 
-\half t_{\rm m}^2 f_2(t) K \Big)\hat{P}_1 
\, +\,  t_{\rm m} f_1(t) 
\sqrt{2\tr\big[\hat\Gamma\Hsys[\Hsys,\hat\Gamma]_-\big]}\,\hat{P}_2.
\end{eqnarray}
The only $f_2(t)$-term that we keep is in the prefactor
on $\hat{P}_1$, for compactness we define 
$K\equiv \tr [\hat\Gamma 
[\hat{\cal H}_{\rm sys},[\hat{\cal H}_{\rm sys},\hat{\Gamma}]_- ]_- ]  
/\sqrt{\tr [\hat\Gamma^2]}$. 
This term gives a ${\cal O}[t_{\rm m}^2]$-term in the final result, 
while other such $f_2(t)$-terms give at worst 
a ${\cal O}[t_{\rm m}^3]$-term. 
From Eq.~(\ref{eq:bs}), we get
\numparts
\begin{eqnarray}
b_+ &=& t_{\rm m}\, f_1(t) 
\,\sqrt{2 \tr \big[ \hat{\Gamma}^2 \big] 
\tr \big[ \hat{\Gamma}\Hsys[\Hsys,\hat{\Gamma}]_- \big] },
\label{eq:bplus-short}
\\
b_z &=& {\rm Re}[f_0(t)]\, \tr \big[ \hat{\Gamma}^2 \big]
- \half t_{\rm m}^2 {\rm Re}[f_2(t)] K \sqrt{\tr \big[ \hat{\Gamma}^2 \big]}.
\label{eq:bz-short}
\end{eqnarray}
\endnumparts
Since $\tr \big[ \hat{\Gamma}\Hsys[\Hsys,\hat{\Gamma}]_- \big]
\lesssim \Delta_{\rm sys}^2\tr \big[ \hat{\Gamma}^2] $,
we have $|b_+| \sim  (\Delta_{\rm sys}t_{\rm m}) b_z$ where   
$\Delta_{\rm sys}t_{\rm m}\ll 1$.
The terms that we dropped
only give contributions of order 
$(\Delta_{\rm sys}t_{\rm m})^2 b_z$.

To zeroth order in $t_{\rm m}$ we recover Lindblad's result, 
Eqs.~(\ref{eq:Lindbladeqn-a},\ref{eq:Lindbladeqn-b}), 
with only one non-zero coupling constant $\lambda_1= 2 {\rm Re}[f_0] 
\tr \big[ \hat{\Gamma}^2 \big] > 0$,
associated with the operator, 
$\hat{L}_1= (\tr[\hat\Gamma^2])^{-1/2}\hat{\Gamma}$.
However to first order in $t_{\rm m}$, we have
Eq.~(\ref{eq:Lindbladeqn-a}) 
with two non-zero coupling constant $\lambda_1$ and $\lambda_2$; the latter of
which is negative (even for infinitesimal $t_{\rm m}$).
We use this model 
to explore the
contradiction between Bloch-Redfield and Lindblad.

\subsection{Environment with a nearly 
white-noise spectrum}
\label{sect:white}

Here we consider an environment with a nearly white-noise spectrum 
of excitations (a very wide Lorentzian),
at extremely high temperature, 
$k_{\rm B}T \gg \om_{\rm m}$,
so
\begin{eqnarray}
S(\om) = S_0\, {\om_{\rm m}^2 \over \om_{\rm m}^2 + \om^2}, 
\qquad 
A(\om) = {\om_{\rm m} S_0\over 2k_{\rm B}T}\,  {\om_{\rm m} \om \over \om_{\rm m}^2 + \om^2}, 
\end{eqnarray}
where $A(\om)$ is given by the
result below Eq.~(\ref{eq:A-omega}).
The high-energy cut-off, $\om_{\rm m}$, 
plays the role of the inverse memory time, $t_{\rm m}^{-1}$,
so for nearly white-noise we need it to be 
much larger than the largest system energy scale, $\Delta_{\rm sys}$.
Then Eq.~(\ref{eq:f_q-S-and-A}) gives
\numparts
\begin{eqnarray}
\fl
f_0(t) &=& \half S_0 
\ \big( 1+\rmi (2k_{\rm B}T t_{\rm m})^{-1}\big) 
\ \big(1-\exp[-t/t_{\rm m}]\big),
\label{eq:f0-white}
\\
\fl
f_1(t)  &=& \half S_0 
\ \big(1+\rmi (2k_{\rm B}T t_{\rm m})^{-1}\big)
\ \big(1-(1+t/t_{\rm m}) \exp[-t/t_{\rm m}]\big),
\label{eq:f1-white}
\\
\fl
f_2(t)  &=& S_0 
\ \big(1+\rmi (2k_{\rm B}T t_{\rm m})^{-1}\big)
\ \big(1-(1+t/t_{\rm m}+ (t/t_{\rm m})^2/2) \exp[-t/t_{\rm m}]\big),
\label{eq:f2-white}
\end{eqnarray}
\endnumparts
where we evaluated the $\om$-integrals using complex analysis
(by pushing the contours into the upper-half plane, one finds that the
results are due to the pole at $\om=\rmi \om_{\rm m}$).
Both $f_0(t)$ and $f_1(t)$ go exponentially to
their long-time limit ($t\gg t_{\rm m}$), with the rate given by the 
memory time, $t_{\rm m}$.  
When $t/t_{\rm m} \gg 1$ we have $f_1(t)/f_0(t) \simeq 1$,
while
when $t/t_{\rm m} \ll 1$ we have $f_1(t)/f_0(t) \simeq t/t_{\rm m}$.
For such an environment, the Bloch-Redfield
equation is valid for $S_0t_{\rm m} \ll 1$,
see \ref{append:Dyson->Bloch}.

%-------------------------------------------------------------------------
\section{Positivity as a constraint on a two-level system's purity}
\label{sect:purity}

To ensure that there is no basis in which the density matrix
has negative probabilities (i.e.~no possible measurement will return an 
unphysical probability)
it is sufficient and necessary that the density matrix's
eigenvalues, $\{\Lambda_k\}$, satisfy $0\leq \Lambda_k \leq 1$ for all $k$.
To see this, consider an arbitrary basis which is 
related to the eigenbasis by the
unitary transformation, ${\cal U}$. In this basis
all probabilities are given by
$\hat \rho_{ii} = \sum_k |\hat{\cal U}_{ik}|^2 \Lambda_k$, 
where the unitarity of
$\hat{\cal U}$ guarantees that $\sum_k |\hat{\cal U}_{ik}|^2=1$.
Thus if $0\leq \Lambda_k \leq 1$ for all $k$, then 
probabilities  in this arbitrary basis,
satisfy $0\leq \hat\rho_{ii} \leq 1$ for all $i$.

A two-level system is special because the eigenvalues of its density matrix
are defined by a single parameter, $s$ (remember that the sum of the
eigenvalues must be one).  
The most general two-by-two density-matrix is of the form
$\hat \rho = \half (\hat\sigma_0
+ s_x \hat\sigma_x + s_y \hat\sigma_y + s_z \hat\sigma_z)$,
where $\hat\sigma_{x,y,z}$ are the Pauli matrices, and 
$s_{x,y,z}$ are real numbers, when diagonalized
it takes the form $\hat{\rho}_{\rm d}= \half (\hat\sigma_0
+ s \hat\sigma_z)$ with the single parameter $s^2 = s_x^2+s_y^2+s_z^2$.
Thus to ensure positivity we require that $-1\leq s \leq 1$.
The purity of $\hat \rho$ is
$P= \tr [\hat{\rho}^2] =\half(1+s_x^2+s_y^2+s_z^2)$, thus ensuring
positivity is equivalent to ensuring that $P\leq 1$. 
This is not the case for systems with more than two levels
\cite{footnote:three-levels}.

Finally it is worth noting that Eq.~(\ref{eq:Lindbladeqn-a}) leads to
\begin{eqnarray}
{\rmd P \over \rmd t} 
= 2 \tr \left[\rhosys(t) {\rmd \rhosys(t) \over \rmd t}\right]
= -2 \sum_{n=0}^{N^2-1} \lambda_n 
\tr \Big[ \hat{L}_n^\dagger \big[\hat{L}_n,\hat\rho(t) \big]_-\,\hat\rho(t) \Big].
\label{eq:dP/dt-general}
\end{eqnarray}

%-------------------------------------------------------------------------
\section{Two-level system with 
nearly white-noise: proving positivity}
\label{sect:2level}

%%%%%%%%%%%%%%%%%%%%%%%%%%%%%%%%%%%%%%%%%%%%
%%%%%% figure goes here %%%%%%
%%%%%%%%%%%%%%%%%%%%%%%%%%%%%%%%%%%%%%%%%%%%
\begin{figure}
\hskip 0.2 \textwidth 
\includegraphics[width=0.7\textwidth]{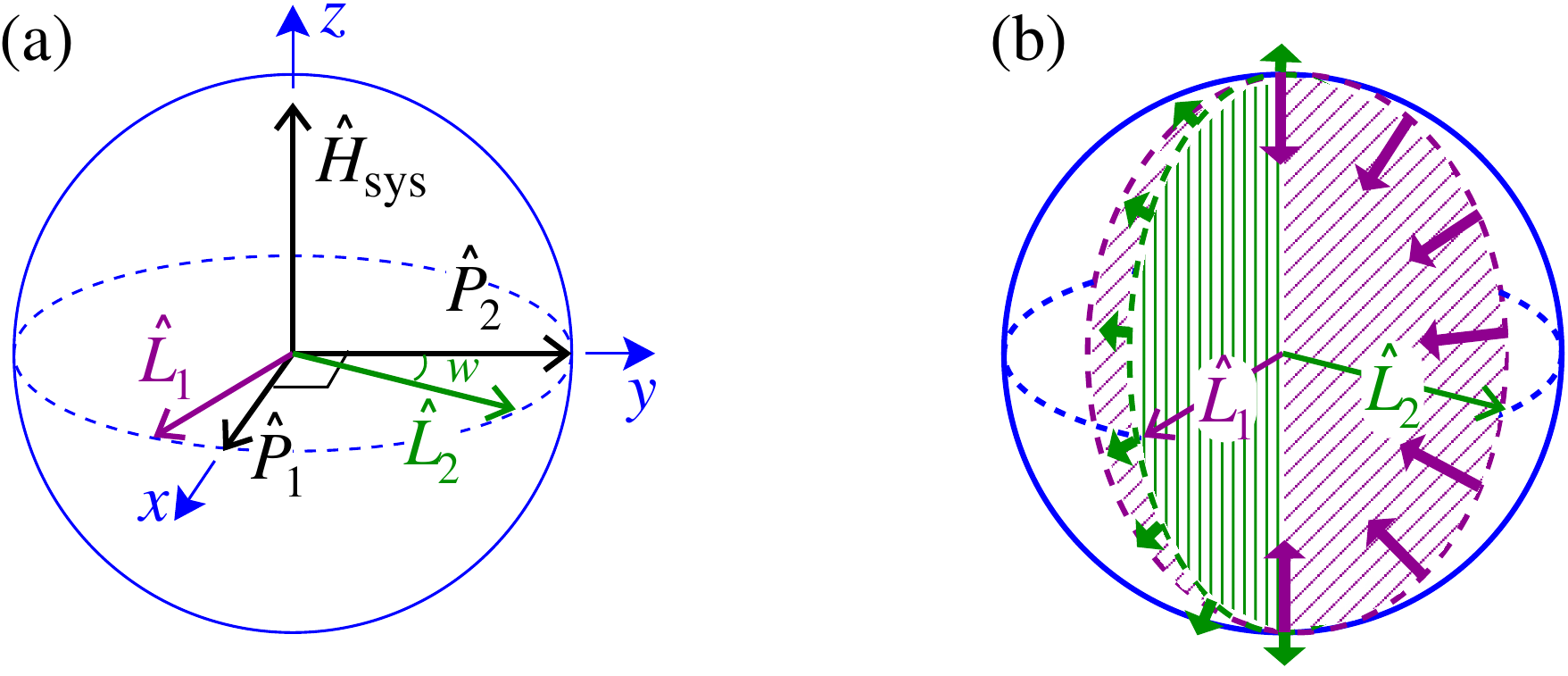}
\caption{
A sketch of the Bloch sphere, 
for the situation discussed in Section~\ref{sect:2level}.
In (a) we show the axes associated
with $\hat{\cal H}_{\rm sys}$, $\hat{P}_{1,2}$ and $\hat{L}_{1,2}$.
We show $\hat{L}_{1,2}$ for a case where they are Hermitian ($w$ is real), 
as only Hermitian operators are associated with axes in the Bloch sphere.
In (b) we sketch the effect of the $\hat{L}_{1,2}$-terms on the evolution
of the Bloch vector which represents the density-matrix,
${\bf r}_{\rm Bloch} = (2{\rm Re}[\rho_{12}], -2{\rm Im}[\rho_{12}],
\rho_{11}-\rho_{22})$.  The $\hat{L}_1$-term {\it reduces} the
magnitude of the vector in the plane perpendicular to $\hat{L}_1$
(diagonal cross-hatching) at a rate given by $\lambda_1$.
The $\hat{L}_2$-term {\it increases} the
magnitude of the vector in the plane perpendicular to $\hat{L}_2$
(vertical cross-hatching)
at a rate given by $|\lambda_2| \ll |\lambda_1|$.
} \label{Fig:Bloch}
\end{figure}
%%%%%%%%%%%%%%%%%%%%%%%%%%%%%%%%%%%%%%%%%%%%

We now consider a two-level system
with ${\cal H}_{\rm sys} = -\half B\hat{\sigma}_z$,
coupled to an environment via $\hat{\Gamma}= \hat{\sigma}_x$.
Then 
Eq.~(\ref{eq:def-Xi}) gives
$\hat\Xi = \int_0^t \rmd \tau \al(\tau) 
\big[\hat{\sigma}_x\cos B\tau - \hat{\sigma}_y \sin B\tau \big]$,
so
$\hat{P}_1 = \hat\sigma_x/\sqrt{2}$ and  
$\hat{P}_2 = \hat\sigma_y/\sqrt{2}$. 
For an environment with a short memory time,
Eqs.~(\ref{eq:bplus-short},\ref{eq:bz-short}) give
$b_+ = 2Bt_{\rm m}\, f_1(t)$
and  
$b_z = 2{\rm Re}[f_0(t)] -(Bt_{\rm m})^2 {\rm Re}[f_2(t)]$.
Thus to second-order in $Bt_{\rm m}$, 
Eq.~(\ref{eq:lambda-and-U}) 
gives
\begin{eqnarray}
\lambda_1 &=& 4{\rm Re}[f_0(t)] 
+ (Bt_{\rm m})^2\left({|f_1(t)|^2 \over {\rm Re}[f_0(t)]} - 2{\rm Re}[f_2(t)]
\right) 
\nonumber \\ 
\lambda_2 &=& - {(Bt_{\rm m})^2|f_1(t)|^2 \over {\rm Re}[f_0(t)]}. 
\label{eq:lambdas-2level}
\end{eqnarray}
Defining $w= {Bt_{\rm m} f_1(t) /{\rm Re}[f_0(t)]}$,
the Lindblad operators, $\hat L_{1,2}$,
are given by
\begin{eqnarray}
\fl
\left(\begin{array}{c} \hat{L}_1 \\ \hat{L}_2 \end{array}\right)
=
{\cal U} \left(\begin{array}{c} \hat{P}_1 \\ \hat{P}_2 \end{array}\right)
= {1 \over \sqrt{2}}
\left(\begin{array}{cc} 
1 - {1\over 8}|w|^2 & w 
\\
w^* & -1 + {1\over 8}|w|^2
\end{array} \right) \left(\begin{array}{c} \hat{\sigma}_x \\ \hat{\sigma}_y \end{array}\right). 
\end{eqnarray}
Here we give ${\cal U}$ to first order in $Bt_{\rm m}$,
but keep the higher order terms necessary to ensure 
${\cal U}^\dagger{\cal U}=1$.
Note that $\hat L_1$ and $\hat L_2$
are not Hermitian unless $f_1(t)$ is real.

We take the $T\to \infty$ limit of the nearly white-noise in 
Section~\ref{sect:white} so ${\rm Im}[f_1(t)]=0$, 
then $w$ is real and gives
the angle marked in Fig.~\ref{Fig:Bloch}a.
Defining the $x',y'$-axes such that
$\hat{L}_1= \hat{\sigma}_{x'}/\sqrt{2}$
and  $\hat{L}_2= \hat{\sigma}_{y'}/\sqrt{2}$,
the Bloch-Redfield equation reduces to
\begin{eqnarray}
\fl
{\rmd \over \rmd t} \rhosys (t)
&=& -\rmi [\Hsys',\rhosys (t)]
- 2\lambda_1
\Big(\rhosys(t) -\hat{\sigma}_{x'} \rhosys(t) \hat{\sigma}_{x'}\Big)
- 2\lambda_2
\Big(\rhosys(t) -\hat{\sigma}_{y'} \rhosys(t) \hat{\sigma}_{y'}\Big).
\label{eq:BlochRed-two-level}
\end{eqnarray}
The coupling constants, $\lambda_1,\lambda_2$, are given by 
Eqs.~(\ref{eq:f0-white},\ref{eq:f1-white},\ref{eq:lambdas-2level}) 
with $T\to \infty$, so
\begin{eqnarray}
\lambda_1 &=& 2 S_0(1-\e^{-t/t_{\rm m}}) + {\cal O}[(Bt_{\rm m})^2],
\nonumber \\
\lambda_2 &=& -(Bt_{\rm m})^2 S_0 
{(1-(1\! +\! t/t_{\rm m}) \e^{-t/t_{\rm m}})^2 \over 2(1-\e^{-t/t_{\rm m}})}.
\label{eq:lambda12-white}
\end{eqnarray}
Substituting these results into Eq.~(\ref{eq:dP/dt-general}),
and writing $-\lambda_2$ as $+|\lambda_2|$ to emphasis
that it tends to increase the purity, 
we get 
\begin{eqnarray}
{\rmd P \over \rmd t} 
= -\lambda_1 \tr \big[\rhosys^2(t) - \big(\sigxp \rhosys(t)\big)^2  \big]
+|\lambda_2| \tr \big[\rhosys^2(t) - \big(\sigyp \rhosys(t)\big)^2 \big].
\label{eq:dP/dt-2level}
\end{eqnarray} 

\subsection{Positivity at short times 
(times of order the memory time)}
\label{sect:short-times}

For times, $t$, much less than $S_0^{-1}$ we can get the purity
to first order in $S_0$, by integrating Eq.~(\ref{eq:dP/dt-2level})
with $\rhosys(t)$ replaced by its value to zeroth order
in $S_0$
\begin{eqnarray}
\fl
\rhosys^{(0)}(t) = \half \Big[1+ 
(s_x\cos Bt +s_y\sin Bt) \sigxp + (s_y\cos Bt -s_x\sin Bt)\sigyp 
+ s_z\sigz \Big],
\label{eq:rhosys-shorttime}
\end{eqnarray}
where the constants $(s_x,s_y,s_z)$ define $\rhosys(t=0)$.  
Note that we have used the fact that to zeroth order in $S_0$ 
we have $\Hsys'= \Hsys = -\half B \sigz$.
Then Eq.~(\ref{eq:dP/dt-2level}) becomes
\begin{eqnarray}
\fl
{\rmd P(t) \over \rmd t} 
= -\lambda_1 \big[ (s_y \cos Bt - s_x \sin Bt)^2 + s_z^2 \big]
+|\lambda_2| \big[ (s_x \cos Bt + s_y \sin Bt)^2 + s_z^2 \big]
\nonumber \\
\qquad \qquad \qquad \qquad \qquad \qquad 
+ {\cal O}[S_0^2t].
\label{eq:dp/dt-approx1}
\end{eqnarray} 
As $t_{\rm m} \ll B^{-1}$,
we can restrict ourselves to times $t \ll B^{-1},S_0^{-1}$
(and hence expand in powers of $Bt$ and $S_0t$),
and still study the dynamics up to times $\gg t_{\rm m}$.
The problematic coupling constant, $\lambda_2$, is ${\cal O}[B^2]$,
so we must expand the right-hand-side
of Eq.~(\ref{eq:dp/dt-approx1}) to ${\cal O}[B^2]$, to see the effect of
$\lambda_2$ on the dynamics.
After this expansion in $B$, we expand the purity about $P(0)=1$.
So $P(t) = 1+ \int_0^t \rmd t'  (\rmd P(t')/\rmd t')$ gives 
\begin{eqnarray}
\fl
 P(t) = 
1 - 2S_0t_{\rm m} 
\Big[ s_z^2\,I_{z}
+s_y^2\,I_{y} 
-2s_xs_y\,Bt_{\rm m}\, I_{xy} 
+ s_x^2 (Bt_{\rm m})^2I_{x}
\Big]
+ {\cal O}[S_0^2t,S_0B^3t^3]
\label{eq:dp/dt-approx2}
\end{eqnarray} 
where $I_z$, $I_y$, $I_{xy}$ and $I_x$ are the following functions of 
$t/t_{\rm m}$,
\numparts
\begin{eqnarray}
\fl
I_{z} (t/t_{\rm m}) 
\equiv &\int_0^t\rmd t'{\lambda_1(t')+\lambda_2(t') \over 2S_0t_{\rm m}}&
\ \simeq \ 
\int_0^{t/t_{\rm m}} \rmd \nu \,(1-\e^{-\nu}),  
\label{eq:Iz}
\\
\fl
I_{y} (t/t_{\rm m}) 
\equiv &\int_0^t \rmd t' {(1-(Bt')^2) \lambda_1(t')\over 2S_0t_{\rm m}}&
\ \simeq\ \int_0^{t/t_{\rm m}} \rmd \nu \,(1-\e^{-\nu}), 
\\
\fl
I_{xy} (t/t_{\rm m}) 
\equiv &\int_0^t\rmd t' {t'\lambda_1(t') \over 2S_0t_{\rm m}^2}&
\simeq \ 
\int_0^{t/t_{\rm m}} \rmd \nu \, \nu(1-\e^{-\nu}),
\\
\fl
I_{x} (t/t_{\rm m}) \equiv 
&\int_0^t\rmd t'{(Bt')^2\lambda_1(t') +\lambda_2(t')\over 2S_0B^2t_{\rm m}^3}&
\nonumber \\
& & \fl \hskip -5mm
\ \simeq\ \int_0^{t/t_{\rm m}} \rmd \nu \Bigg[ 
\nu^2(1-\e^{-\nu}) - { [1-(1+\nu)\e^{-\nu}]^2 \over 4(1-\e^{-\nu})} \Bigg],
\label{eq:Ix}
\end{eqnarray}
\endnumparts
where $\nu = t'/t_{\rm m}$.
The ``$\simeq$'' indicates that we keep only the leading order
in $(Bt_{\rm m})$ in each term, this will be sufficient for our purposes.

%%%%%%%%%%%%%%%%%%%%%%%%%%%%%%%%%%%%%%%%%%%%
%%%%%% figure goes here %%%%%%
%%%%%%%%%%%%%%%%%%%%%%%%%%%%%%%%%%%%%%%%%%%%
\begin{figure}
\hskip 0.15 \textwidth
\includegraphics[width=0.8\textwidth]{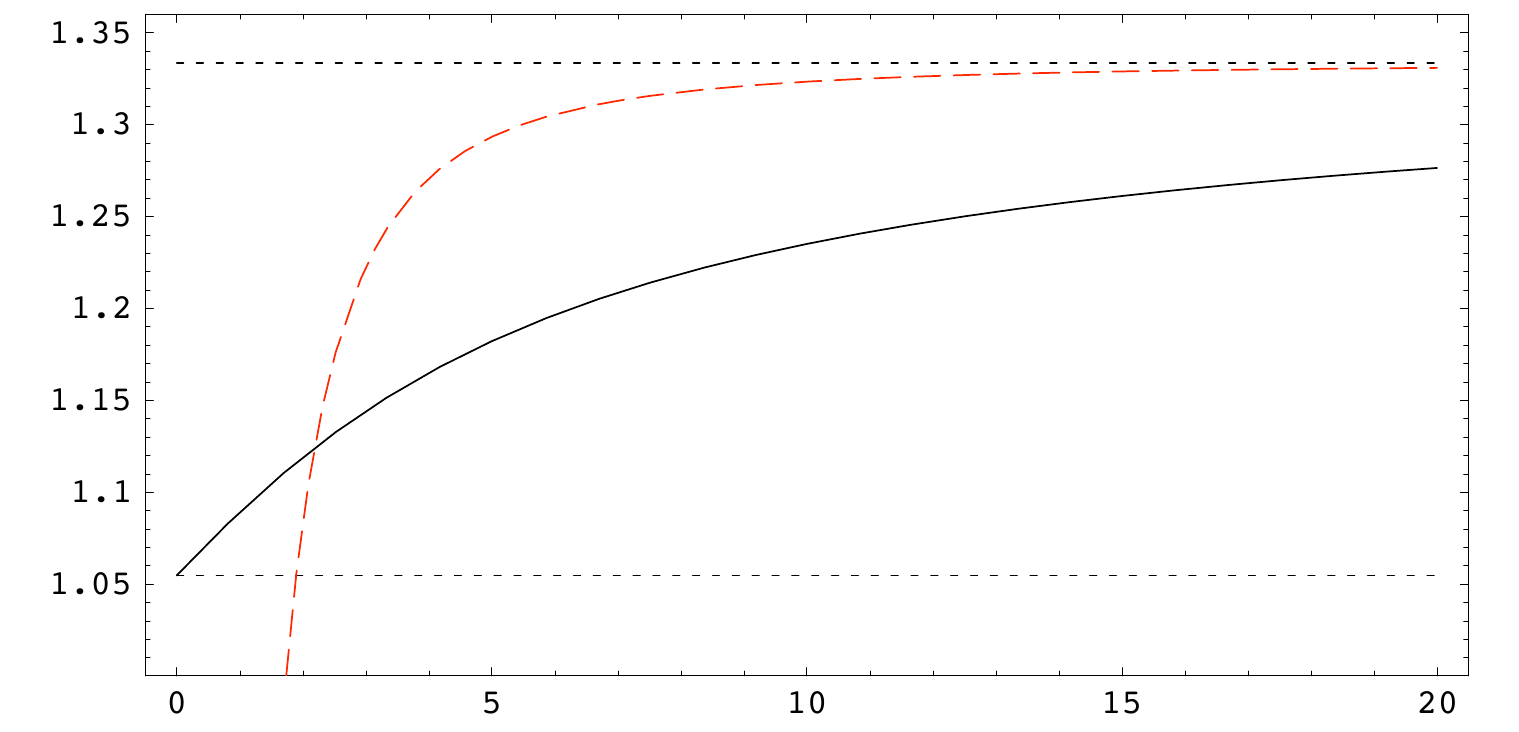}
\caption{
Plots of $I_yI_x/I_{xy}^2$ (solid curve) and 
$I_y^{(\infty)}I_x^{(\infty)}/(I_{xy}^{(\infty)})^2$ (dashed curve) as
functions of $t/t_{\rm m}$.
The two horizontal lines are the two extrema of  $I_yI_x/I_{xy}^2$;
its small $t$ limit of $135/128$ and 
its large $t$ limit of $4/3$.
The crucial point is that  $I_yI_x/I_{xy}^2 > 1$ for all $t\geq 0$.
This is not the case for
$I_y^{(\infty)}I_x^{(\infty)}/(I_{xy}^{(\infty)})^2$,
which one would get if one mistakenly assumed time-independent
coupling constants $\lambda_1(\infty),\lambda_2(\infty)$; 
this is less than one 
for all $t/t_{\rm m} <\sqrt{3}$ and goes to $-\infty$ at $t=0$.
} \label{Fig:plot}
\end{figure}
%%%%%%%%%%%%%%%%%%%%%%%%%%%%%%%%%%%%%%%%%%%%

To show that $P(t)$ does not exceed one 
(in the range of $t$ for which Eq.~(\ref{eq:dp/dt-approx2}) is valid), 
we show that the square-bracket in Eq.~(\ref{eq:dp/dt-approx2}) 
is never negative.
Writing the square-bracket as
$\big[I_{z} s_z^2+I_{y}(s_y -s_x(Bt_{\rm m})I_{xy}/I_{y})^2 
+s_x^2(Bt_{\rm m})^2 
(I_{x}-I_{xy}^2/I_{y})\big]$,
we see that there are three terms;
the first two are always positive (but will be small for spins starting
close to the $x'$-axis, i.e. 
$s_y,s_z\ll 1$),
the third term is positive if $I_x > I_{xy}^2/I_{y}$.
Thus 
we must show that  $I_yI_x/I_{xy}^2\geq 1$.
For $t  \ll t_{\rm m}$,
\begin{eqnarray}
I_y \to \half (t/t_{\rm m})^2 \qquad
I_{xy} \to \third (t/t_{\rm m})^3 \qquad
I_x \to {\textstyle \frac{15}{64}}  (t/t_{\rm m})^4,
\end{eqnarray}
and for $t \gg t_{\rm m}$,
\begin{eqnarray}
I_y \to  t/t_{\rm m} \qquad
I_{xy} \to  \half (t/t_{\rm m})^2 \qquad
I_x \to \third (t/t_{\rm m})^3.
\label{eq:Is-at-infinity}
\end{eqnarray}
Thus for $t  \ll t_{\rm m}$ we have $I_yI_x/I_{xy}^2\to 135/128$,
while for $t \gg t_{\rm m}$ we have $I_yI_x/I_{xy}^2 \to 4/3$.
For finite $t$ we see that $I_yI_x/I_{xy}^2$ is a monotonic function
of $t$ which goes from $135/128$ to $4/3$
(see Fig.~\ref{Fig:plot}), thus it is always greater than one.
This means that $P(t)\leq 1$ for {\it all} times much less than $S_0^{-1}$
(including times greater than $t_{\rm m}$).

For $t_{\rm m} \ll t \ll B^{-1},S_0^{-1}$, 
the purity 
$P(t)=1- 2S_0t \big[ s_z^2 +\big(s_y- \half s_xBt\big)^2 
+ {\textstyle \frac{1}{12}}  (s_xBt)^2  \big]$.
By checking all pure initial system states 
(state with $s_x^2+s_y^2+s_z^2=1$)
we see that this is maximal for 
$s_x=\pm [1+(Bt)^2/8] +{\cal O}[B^3t^3]$, 
$s_y = \pm \half Bt$ and $s_z=0$ (where $s_x$ and $s_y$ have the same sign)
\cite{footnote:polar}. Hence 
\begin{eqnarray}
P(t_{\rm m} \ll t \ll B^{-1},S_0^{-1}) 
\ \leq \ 1- {\textstyle \frac{1}{6}}S_0B^2t^3.
\label{eq:P-bound-short-time}
\end{eqnarray}
This upper-bound on the purity
will be crucial for our proof (in section~\ref{sect:long-times}) 
that the purity does not exceed one
at longer times.

{\it If
we had made the usual assumption} that one can replace
$\lambda_1(t)$ and $\lambda_2(t)$ with $\lambda_1(t=\infty)$ 
and $\lambda_2(t=\infty)$ for all $t$,
then we would have Eqs.~(\ref{eq:Iz}-\ref{eq:Ix}) 
with $I_y,I_{xy},I_x$ 
replaced 
by 
$I_y^{(\infty)}= \int_0^{t/t_{\rm m}} \rmd \nu$,
$I_{xy}^{(\infty)} = \int_0^{t/t_{\rm m}} \rmd \nu \, \nu$,
and 
$I_x^{(\infty)} = \int_0^{t/t_{\rm m}} \rmd \nu \left[\nu^2 - 1/4 \right]$.
We plot $I_y^{(\infty)}I_x^{(\infty)}/(I_{xy}^{(\infty)})^2$
in Fig.~\ref{Fig:plot}, and see that it goes to $-\infty$ as
$t/t_{\rm m} \to 0$.
Thus such a mistaken assumption would have led us to conclude 
(as other have) that $P$ can become bigger than one 
(at times $\lesssim t_{\rm m}$). 
The mistake is most clearly illustrated by looking at 
Eq.~(\ref{eq:dp/dt-approx2}) with $s_x=1$ and $s_y=s_z=0$, then
using $I_x^{(\infty)}$ in place of $I_x$ would lead one to think
that for $t \ll t_{\rm m}$, the purity would be $1+\half S_0t$
when the correct expression shows it is
$1- \frac{2}{3}S_0t_{\rm m}(t/t_{\rm m})^3$.
Thus it is only by keeping the time-dependence of the
coupling constants, that we can show that the purity 
cannot exceed one 
for all times $\ll S_0^{-1}$ (including times greater than $t_{\rm m}$).

\subsection{Positivity at long times (times of order and greater than 
$S_0^{-1}$)}
\label{sect:long-times}

We now turn to the evolution of the purity
at all times much greater than $t_{\rm m}$ (the long-time regime in 
Fig.~\ref{Fig:timescales}).
For times of order and greater than $S_0^{-1}$ 
we need the full Bloch-Redfield equation, Eq.~(\ref{eq:BlochRed-two-level}),
not just the short time expansion of it.
Since $t \gg t_{\rm m}$, the
coupling constants have saturated at their long time limits;
$\lambda_1 = 2 S_0$ and $\lambda_2 = -2 S_0 (Bt_{\rm m}/2)^2$.
Then Eq.~(\ref{eq:dP/dt-2level}) reduces to
\begin{eqnarray}
{\rmd P(t) \over \rmd t} 
= -2 S_0 \big[(1-(Bt_{\rm m}/2)^2)s_z^2(t) + s_y^2(t) 
- (Bt_{\rm m}/2)^2s_x^2(t) \big].
\label{eq:dP/dt-longtimes}
\end{eqnarray} 
Since $Bt_{\rm m} \ll 1$, we can see that
$P(t\gg t_{\rm m})$ decays for nearly all 
$s_{x,y,z}(t)$. However the purity may grow if  
$s_y(t) \sim s_z(t) \sim {\cal O}[(Bt_{\rm m})^2]$; 
then the purity might exceed one (particularly if $s_x$ is close to one).

To see if the purity can exceed one, 
we expand the evolution about the time $t_0$, where
we choose $t_0$ such that $s_y(t_0)=0$.
We then perform the same expansion about $t=t_0$ here as we performed
about $t=0$ in Section~\ref{sect:short-times}.
Hence on the right-hand-side of Eq.~(\ref{eq:dP/dt-longtimes})
we make the substitution
$s_x(t_0+\tau) = s_x' \cos B\tau$,
$s_y(t_0+\tau) = s_x' \sin B\tau$,
$s_z(t_0+\tau) = s_z'$,
where we define $s_x'=s_x(t_0)$ and $s_z'=s_z(t_0)$ 
(remember that $t_0$ is chosen such that $s_y(t_0)=0$).
This substitution is good for all $\tau \ll S_0^{-1}$,
After the substitution we expand the right-hand-side of 
Eq.~(\ref{eq:dP/dt-longtimes}) up to second order in $B\tau$.
Thus for $\tau \ll B^{-1},S_0^{-1}$,
\begin{eqnarray}
\fl
\left. {\rmd P(t) \over \rmd t}\right|_{t=t_0+\tau} 
= -2 S_0 \big[(1-(Bt_{\rm m}/2)^2){s_z'}^2 + (s_x' B\tau)^2 
- (s_x'Bt_{\rm m}/2)^2 \big].
\label{eq:dP/dt-longtimes2}
\end{eqnarray} 
From this we see that the purity can only increase during
a time-window  where $|\tau| < [(t_{\rm m}/2)^2 - (s_z'/s_x'B)^2]^{1/2}$
(neglecting a term that is higher order in $Bt_{\rm m}$).
The maximum possible  time for this growth is $t_{\rm m}$
(i.e. when $s_z'=0$, $P$ grows during the time-window
from $\tau=-\half t_{\rm m}$ to $\tau=\half t_{\rm m}$).
Thus the assumption that $\tau \ll B^{-1},S_0^{-1}$
is fulfilled for all $\tau$ at which the purity is growing.

At this point it is sufficient to make a gross over-estimate of
the amount by which the purity can grow.
If we assumed that the purity grows during the entire time-window
$-\half t_{\rm m} \leq \tau \leq \half t_{\rm m}$
at the maximal possible rate 
(i.e. the rate at $\tau=0$ when $s_z'=0$ and $s_x=1$), 
then during this time-window it would grow by $\half S_0B^2t_{\rm m}^3$.
If we define $\Delta P$ as the true increase of the purity in
the time-window where it grows,
the over-estimate enables us to put the following 
upper-bound; 
\begin{eqnarray}
\Delta P <\half S_0B^2t_{\rm m}^3.
\end{eqnarray}
Comparing this with the upper-bound on the purity in 
Eq.~(\ref{eq:P-bound-short-time})
with $t \gg t_{\rm m}$ (but $t \ll S_0^{-1}$), 
we see that 
increasing the purity by $\Delta P$
cannot cause it to exceed one.
The short- and long-time regimes overlap (see Fig.~\ref{Fig:timescales}), 
so by showing that $P\leq1$ in both regimes we have shown 
positivity for all $t>0$.

%-------------------------------------------------------------------------
\section{Conclusions}
\label{sect:concl}

The Bloch-Redfield master equation 
for an arbitrary system can be written
in the form of a Lindblad master equation, Eq.~(\ref{eq:Lindbladeqn-a}).
Only by setting the memory time equal to zero (strictly Markovian evolution)
do we recover Lindblad's result with coupling constants, $\{\lambda_n\}$, 
which are time-independent and positive, Eq.~(\ref{eq:Lindbladeqn-b}).

For finite memory times,
the Bloch-Redfield master equation can still be cast in the form of
Eq.~(\ref{eq:Lindbladeqn-a}),
but its do not satisfy Eq.~(\ref{eq:Lindbladeqn-b}).
However, the parameters are time-dependent 
which means that the semigroup property is absent, 
and so Lindblad's requirements are {\it inapplicable}.
We show analytically for a particular model 
(a two-level system coupled to a high-temperature environment
with a memory time much less than system timescales)
that the master equation preserves {\it positivity} 
if and only if we keep the time-dependence of the parameters.
Further, it turns out that {\it positivity} and {\it complete positivity} 
are equivalent for this particular model \cite{Hall08}.

It is remarkable that our result only coincides with
Lindblad's for strictly zero memory time, $t_{\rm m}=0$.
If we take the limit $t_{\rm m}\to 0$, we find that
one coupling constant tends to zero from {\it below}.
Further, we argue (see the appendix) that the Bloch-Redfield equations
become exact in this limit.
Thus even for infinitesimal $t_{\rm m}$, 
one coupling constant is negative.
Positivity (and hence complete positivity)
is none-the-less preserved by the
time-dependence of the coupling constants at
times of order the infinitesimal time $t_{\rm m}$.

We wonder if an analysis of the time-dependent parameters of 
an arbitrary Bloch-Redfield master equation would show positivity,
or even complete positivity.
If this could be proven, one could argue that
the Bloch-Redfield master equation contains both the
Lindblad equation and finite memory-time corrections to it.

%-------------------------------------------------------------------------
\section{Acknowledgments}

This work was stimulated by conversations with 
J.~Siewert, Y.~Gefen and S.~Stenholm,
at the ``Workshop on entanglement, decoherence and geometric phases 
in complex systems'', Abdus Salam ICTP, 2004.
I am extremely grateful to M.~Hall for pointing out a serious mistake in 
the first draft of this manuscript, and for enlightening discussions.
My thanks also go to A.~Shnirman, M.~Clusel
and D.~O'Dell for useful discussions.
The Swiss NSF financed early stages of this work, 
part of which was carried out at the Aspen Centre for Physics.

%-------------------------------------------------------------------------
%-------------------------------------------------------------------------

\appendix

\section{Deriving Bloch-Redfield from a Dyson equation}
\label{append:Dyson->Bloch}

For completeness, 
we sketch the derivation of the 
Bloch-Redfield master equation \cite{Bloch57,Redfield57},
using a common ``modern'' approach \cite{Makhlin-review03}
based on a real-time Dyson equation \cite{Schoeller94}.
The derivation is none-the-less equivalent to Refs.~\cite{Bloch57,Redfield57}.
At $t=0$ the system and environment are in a factorized state
(e.g.~a perfect projective measurement is made on the system at $t=0$). 
The propagator of the system's reduced density matrix is 
${\mathbb K}_{i'j';ij}(t;0) = \tr_{\rm env} \big[ 
\langle i'| \e^{-\rmi \Huniv t} |i\rangle
\hat{\rho}_{\rm env}
\langle j| \e^{\rmi \Huniv t} |j'\rangle
 \big]$, with 
Eq.~(\ref{eq:K}) giving the system's reduced density-matrix at time $t$.
The Dyson equation for ${\mathbb K}(t;t_0)$ 
(treating the system-environment interaction as a perturbation,
which we keep to all orders) is 
\begin{eqnarray}
\fl
{\mathbb K}(t;0)
= {\mathbb K}^{\rm sys}(t;0)
+  \int_{0}^t \rmd t_2 \int_{0}^{t_2} \rmd t_1
{\mathbb K}^{\rm sys}(t;t_2) {\bSigma}(t_2;t_1){\mathbb K}^{\rm sys}(t_1;0),
\label{eq:dyson}
\end{eqnarray}
where ${\mathbb K}(t;t')$ is the propagator including all interactions;
${\mathbb K}^{\rm sys}(t;t')$ is the bare system propagator 
(propagating it only under the Hamiltonian $\Hsys$).
Since there are no interactions after $t=0$ in the first term
and after time $t_2$ in the second term above, we can trace out the 
environment at these times.
Finally
${\bSigma}(t_2;t_1)$ an {\it irreducible} block of the propagator
(with the same tensor structure as 
${\mathbb K}(t;0)$), it is the smallest block for which
the system has interacted with one or more environment excitations.

Taking the time-derivative 
of Eq.~(\ref{eq:dyson}), 
and noting that
$
(\rmd/\rmd t) \int_0^t \rmd t_2 {\mathbb K}^{\rm sys}(t;t_2)
{\mathbb F}(t_2)$
$ 
= {\mathbb F}(t)  + \int_0^t \rmd t_2 {\mathbb E}^{\rm sys}(t) 
{\mathbb K}^{\rm sys}(t;t_2) {\mathbb F}(t_2)$
for any ${\mathbb F}(t_2)$, we get the master equation
\begin{eqnarray}
{\rmd \over \rmd t}{\mathbb K}(t;0) 
&=& -\rmi {\mathbb E}^{\rm sys}(t){\mathbb K}(t;0) 
+ \int_0^t \rmd t_1 {\bSigma}(t;t_1){\mathbb K}(t_1;0).
\label{eq:generalised-master1}
\end{eqnarray}
We have defined 
${\mathbb E}^{\rm sys}_{i'j';ij}= \langle i'|{\cal H}_{\rm sys}|i\rangle 
\langle j|{\cal H}_{\rm sys}|j'\rangle$, then 
$(\rmd/\rmd t){\mathbb K}^{\rm sys} (t;0)
= -\rmi {\mathbb E}^{\rm sys}(t) {\mathbb K}^{\rm sys} (t;0)$.
To clearly see the non-Markovian nature of Eq.~(\ref{eq:generalised-master1})
we can substitute it into Eq.~(\ref{eq:K}) 
which gives
$(\rmd /\rmd t)\rhosys(t) 
= -\rmi [\Hsys(t),\rhosys(t)] 
+ \int_0^t \rmd t_1 {\bSigma}(t;t_1)\rhosys(t_1)$.
This master equation is exact, 
our only assumption was that the system and environment 
were in a factorized state at time $t=0$.
It is formally equivalent to the 
Nakajima-Zwanzig equation \cite{Nakajima58,Zwanzig60}.
However it is of little practical use (giving
no great advantage over standard perturbation theory) unless
the irreducible block, ${\bSigma}(t_2;t_1)$ is reasonably local in time,
i.e. decays on a scale $t_2-t_1\ll t$.
Without approximation we can use
$\rhosys(t)={\mathbb K}(t;t_1) \rhosys(t_1)$ to
write this master equation as
\begin{eqnarray}
{\rmd \over \rmd t}\rhosys(t) 
&=& -\rmi [\Hsys(t),\rhosys(t)] 
+ \int_0^t \rmd t_1 {\bSigma}(t;t_1){\mathbb K}^{-1}(t;t_1) \rhosys(t).
\label{eq:generalised-master3}
\end{eqnarray}
This might ``look'' Markovian, 
but the non-Markovian nature is in the new term
${\mathbb K}^{-1}(t;t_1)$.
Approximations of Eq.~(\ref{eq:generalised-master3})
will give a Bloch-Redfield master equation.

%%%%%%%%%%%%%%%%%%%%%%%%%%%%%%%%%%%%%%%%%%%%
%%%%%% figure goes here %%%%%%
%%%%%%%%%%%%%%%%%%%%%%%%%%%%%%%%%%%%%%%%%%%%
\begin{figure}
\epsfxsize = 1.0 \textwidth 
\hfill
\includegraphics[width=1.0\textwidth]{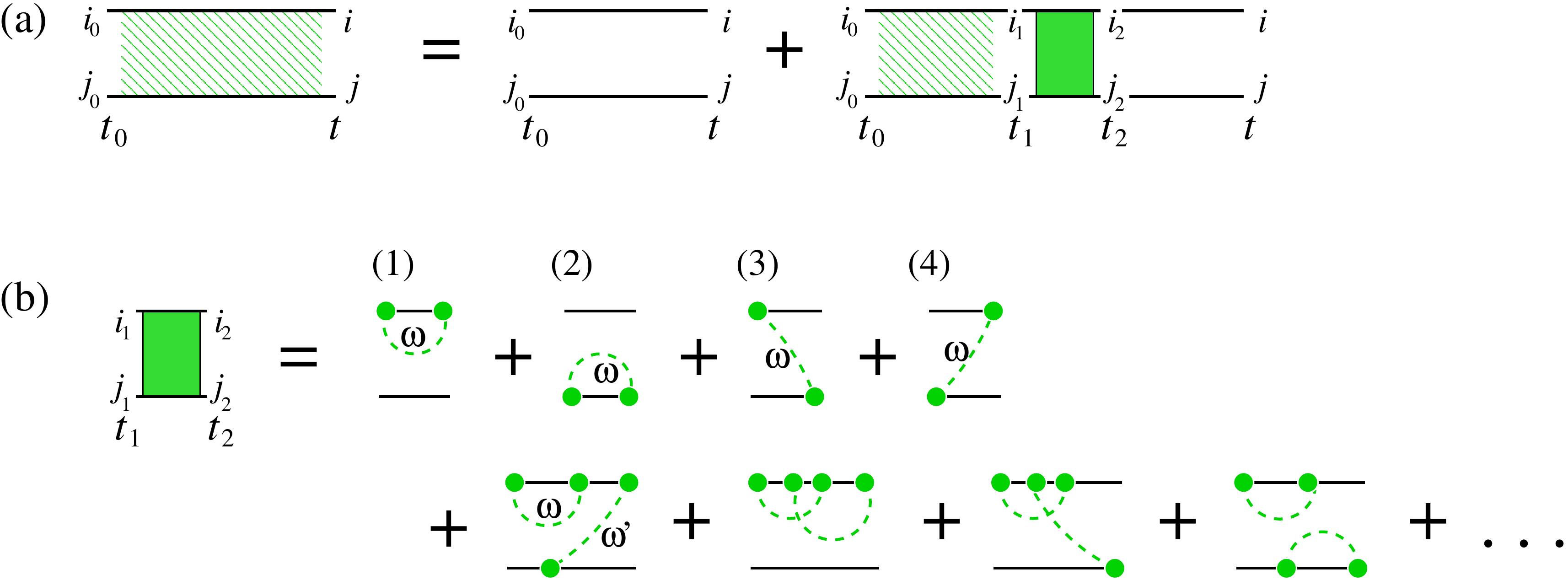}
\caption{(a) Real time Dyson equation for an arbitrary system.
The pair of lines with cross-hatching between them
are the full propagator, ${\mathbb K}(t'';t')$;
the lines without cross-hatching
are the bare system propagator, ${\mathbb K}^{\rm sys}(t'';t')$;
and the lines with solid colour between them
are the irreducible block, ${\bSigma}(t'',t')$.
Internal indices are summed over and
internal times are integrated over 
as in Eq.~(\ref{eq:dyson}).
This drawing of the propagators emphasizes that
only ${\mathbb K}^{\rm sys}_{i'',j'';i'j'}(t'';t')$ 
can be written in the form $A_{i'';i'} \times B_{j'';j'}$.
(b) Some lower-order diagrams for the 
irreducible block, ${\bSigma}(t'',t')$, 
in all cases we integrate $\om$, $\om'$, etc, 
over the spectrum of excitations.
The second-order diagrams are labelled (1) to (4).
} \label{Fig:Dyson}
\end{figure}
%%%%%%%%%%%%%%%%%%%%%%%%%%%%%%%%%%%%%%%%%%%%

\subsection{The Bloch-Redfield equation from a Born approximation}

Here we get
the Bloch-Redfield master equation by making
a Born approximation of the irreducible block ${\bSigma}(t'',t')$
in Eq.~(\ref{eq:generalised-master3}). It involves neglecting all
contributions to ${\bSigma}(t'',t')$ beyond second-order.
Our derivation involves two assumptions which justify the Born approximation
(other derivations may be possible).

Our first assumption is that the environment is large enough to
have a continuous energy-spectrum of excitations 
(although it does not matter if this spectrum is discrete on scales 
$\ll t^{-1}$). 
So for finite relaxation/decoherence rates,
we assume the coupling to each environment excitation is
small enough to be treated only up to second order.
Thus each excitation evolves only under $\Henv$ up to the time of
its (first or second order) interaction with the system.
It then never interacts with the system again, so
we trace it out immediately after the (first or second order) interaction.

Our second assumption is that the environment's initial density-matrix
obeys $[\Henv,\hat\rho_{\rm env}]=0$, as would be the case for
either an eigenstate or any classical mixture of eigenstates of $\Henv$
(such as a thermal state).  
Combining this with our first assumption means that we can treat
$\hat\rho_{\rm env}$ as time-independent.
Then without loss of generality we can make 
${\rm tr}_{\rm env} \big[
\hat{x} \hat{\rho}_{\rm env} \big]=0$,
by moving any constant off-set into the definition of $\Hsys$.
This removes the first order contributions from  
the irreducible block,
${\bSigma}(t'',t')$. 
Thus ${\bSigma}(t'',t')$ becomes the sum of second-order 
(and higher-order) terms sketched in Fig.~\ref{Fig:Dyson}b.
The dotted lines indicate that a given environment excitation
(with energy $\om$) 
has been created by the system-environment interaction.

Treating the integral in Eq.~(\ref{eq:generalised-master3})
to lowest (second) order in $\hat x$, means making a Born approximation
on ${\bSigma}(t'',t')$ and treating ${\mathbb K}^{-1}(t;t_1)$
to zeroth order in $\hat x$ \cite{footnote:difference}.
Hence defining
$\tau= t-t_1$ and 
${\mathbb S} (\tau)= {\bSigma}^{\rm Born}(t;t-\tau)
[{\mathbb K}^{\rm sys}(t;t-\tau)]^{-1}$, we have
\begin{eqnarray}
{\rmd \over \rmd t}\rhosys(t) 
&=& -\rmi [\Hsys(t),\rhosys(t)] 
+ \int_0^t \rmd \tau {\mathbb S}(\tau) \rhosys(t).
\label{eq:markovian-master}
\end{eqnarray}
The four contributions to ${\bSigma}^{\rm Born}(t;t-\tau)$,
labelled (1-4) in  Fig.~\ref{Fig:Dyson}b, give
\numparts
\begin{eqnarray}
{\mathbb S}^{(1)}_{i'j';ij}(\tau) &=& 
\tr_{\rm env} \big[ 
\langle i'|\hat\Gamma \hat{x} \e^{-\rmi \hat{\cal H}_0\tau} 
\hat\Gamma \hat{x} \e^{\rmi \hat{\cal H}_0\tau}| i \rangle 
\rho_{\rm env} 
\langle j| j'\rangle \big], 
\\
{\mathbb S}^{(2)}_{i'j';ij}(\tau) &=& 
\tr_{\rm env} \big[ \langle i'| i \rangle 
\rho_{\rm env} 
\langle j| \e^{-\rmi \hat{\cal H}_0\tau} \hat\Gamma \hat{x} 
\e^{\rmi \hat{\cal H}_0\tau} \hat\Gamma \hat{x}|j'\rangle \big],
\\
{\mathbb S}^{(3)}_{i'j';ij}(\tau) &=& 
\tr_{\rm env} \big[ \langle i'| \e^{-\rmi \hat{\cal H}_0\tau} 
\hat\Gamma \hat{x} \e^{\rmi \hat{\cal H}_0\tau}|i \rangle 
\rho_{\rm env} 
\langle j|\hat\Gamma \hat{x} |j'\rangle \big],
\\
{\mathbb S}^{(4)}_{i'j';ij}(\tau) &=& 
\tr_{\rm env} \big[ \langle i'| \hat\Gamma \hat{x}|i \rangle 
\rho_{\rm env} 
\langle j| \e^{\rmi \hat{\cal H}_0\tau} 
\hat\Gamma \hat{x} \e^{-\rmi \hat{\cal H}_0\tau} |j'\rangle \big],
\end{eqnarray}
\endnumparts
where we define $\hat{\cal H}_0= \Hsys +\hat{\cal H}_{\rm env}$.
We re-write all these contributions in terms of
operators acting to the left and right of the density-matrix, $\rhosys (t)$.
Those interaction on the upper line are to the left of $\rhosys (t)$,
while those on the lower line are to the right.
Thus summing these four terms we get
${\mathbb S}(\tau) \rhosys(t) 
= \tr_{\rm env} \big[\,  
\hat{\Gamma}(0)\hat{x}(0) \hat{\Gamma}(-\tau)\hat{x}(-\tau) 
[\rhosys(t)\otimes \rho_{\rm env}] 
\, +\, [\rhosys(t)\otimes \rho_{\rm env}] 
\hat{\Gamma}(-\tau)\hat{x}(-\tau) \hat{\Gamma}\hat{x}
\, +\, \hat{\Gamma}(-\tau)\hat{x}(-\tau)[\rhosys(t)\otimes \rho_{\rm env}] 
\hat{\Gamma}(0)\hat{x}(0)
\, +\, \hat{\Gamma}(0)\hat{x}(0)[\rhosys(t)\otimes \rho_{\rm env}] 
\hat{\Gamma}(-\tau)\hat{x}(-\tau) 
\big]$, 
where the operators are in the interaction picture,
so
$\hat{\Gamma}(\tau) = \exp[\rmi \Hsys \tau]\hat{\Gamma} \exp[-\rmi \Hsys \tau]$
and
$\hat{x}(\tau) = \exp[\rmi \Henv \tau]\hat{x} \exp[-\rmi \Henv \tau]$.
Substituting this into Eq.~(\ref{eq:markovian-master}) we get the 
Bloch-Redfield master equation that we gave in Section \ref{sect:Bloch}.

Finally, to see {\it when} the Born approximation is justified, we
must estimate the higher-order contributions that we are neglecting.
The higher order contributions to ${\bSigma}(t;t_1)$
take a similar form to the second-order ones, 
but have more factors of $\hat{\Gamma}\hat{x}$ 
acting to the left and right of the density-matrix.   
The times at which these interactions can occur are chosen such that
that ${\bSigma}(t;t_1)$ is irreducible (as discussed above).
A typical fourth order contribution
(those in the second line of Fig.~\ref{Fig:Dyson}b)
will go like $|\hat{\Gamma}\hat{x}|^4t_{\rm m}^3$,
compared with the second-order terms that went like 
$|\hat{\Gamma}\hat{x}|^2t_{\rm m}$.
It is justifiable to neglect the fourth-order 
while keeping the second-order, only if $|\hat{\Gamma}\hat{x}|t_{\rm m} \ll 1$.
Physically the constraint that $|\hat{\Gamma}\hat{x}|t_{\rm m} \ll 1$
means that the Bloch-Redfield master equation applies to situations
where the decay rate of memory effects, $t_{\rm m}^{-1}$, 
is much faster than dissipative 
(relaxation and decoherence) rates $\sim |\hat{\Gamma}\hat{x}|^2t_{\rm m}$.
There is no constraint on the ratio of 
dissipative rates to the system's energy-scales,
so the Bloch-Redfield equation can be applicable to
strong (over-damped) and weak (under-damped) dissipation.

\section{A simple picture of initial-slips}
\label{append:initial-slip}

To understand how {\it initial-slips} work 
\cite{Geigenmuller83,Haake-Lewenstein83,
Haake-Reibold85,Gorini89,Suarez92,Gnutzmann-Haake96,
Gaspard-Nagaoka99,Yu00,Maniscalco03,Cheng05},
it helpful 
to neglect the matrix structure of the master equation. 
Then one has
\begin{eqnarray}
  ({\rm d}/{\rm d} t) \rho(t) = -F(t) \,\rho(t), 
\label{eq:good}
\end{eqnarray}
where $F(t)$ is time-dependent, but saturates at a finite value, $f_\infty$, 
for times greater than the memory time 
($F$ and $\rho$ are now numbers not matrices). 
This is traditionally approximated 
by \cite{book:Cohen-Tannoudji,book:open-quantum}
\begin{eqnarray}
  ({\rm d}/{\rm d}t) \rho(t) = -f_\infty \,\rho(t).
\label{eq:bad}
\end{eqnarray}
Eq.~(\ref{eq:bad}) gives the wrong evolution for any initial condition, 
$\rho(0)$, because it is not justified for times less than the memory time.
However, by multiplying $\rho(0)$ by an {\it initial-slip} one can ensure the 
evolution under the incorrect Eq.~(\ref{eq:bad}) coincides 
with the evolution under the correct Eq.~(\ref{eq:good}) 
for all times much {\it greater} than the memory time.
For the above equations, the initial-slip is simply
$\exp \big[-\int_0^t dt' [F(t') -f_\infty] \big]$.
For $t$ much greater than the memory time, $t_{\rm m}$, 
the initial-slip becomes time-independent
(one can take the integral's upper-limit to $\infty$
since $[F(t')-f_\infty] \sim 0$ for $t'\gg t_{\rm m}$).  
Thus one can take $\rho(0)$, ``slip it'' so that it becomes
$\exp \big[-\int_0^\infty dt' [F(t') - f_\infty] \big] \rho(0)$, 
and use that as the initial 
condition for evolution under the incorrect Eq.~(\ref{eq:bad}).
The resulting $\rho(t)$ coincides with the correct result for all
times much greater than $t_{\rm m}$, but will be absolutely
meaningless for all times of order $t_{\rm m}$. 
Qualitatively the same analysis applies to the full master equation,
but it is complicated by the matrix
structure of the master equation (see e.g.~Ref.~\cite{Gaspard-Nagaoka99}).

The above sketch of the initial-slip method, makes it clear 
that it is {\it not} suited to our analysis of positivity.
The short-time dynamics (on timescales of order the memory time) that it 
generates are {\it fictitious}; 
a sudden initial-slip of the density matrix 
followed by evolution under an incorrect master equation.
Studying the positivity for these fictitious
short-time dynamics tells us nothing about whether the 
true short-time dynamics preserves positivity or not.

%%%% Bibliography %%%%%%%%%%%%%%%%%%%%%%%%%%%%%%%%%%%%%%%%%%%%%%%%%%%%

%%%%%%%%%%%%%%%%%%%%%%%%%%%%%%%%%%%%%%%%%%%%%%%%%%%%%%%%%%%%%%%%%%%%%%%
%%%%%%%%%%%%%%%%%%%%%%%%%%%%%%%%%%%%%%%%%%%%%%%%%%%%%%%%%%%%%%%%%%%%%%%
%%%%%%%%%%%%%%%%%%%%%%%%%%%%%%%%%%%%%%%%%%%%%%%%%%%%%%%%%%%%%%%%%%%%%%%
%%%%%%%%%%%%%%%%%%%%%%%%%%%%%%%%%%%%%%%%%%%%%%%%%%%%%%%%%%%%%%%%%%%%%%%


\begin{thebibliography}{10}
\bibitem{footnote:earlier-arguments-for positivity}
In \cite{Yu00} there is an argument for the preservation of 
positivity
for a damped harmonic oscillator. We believe that this argument is correct
for infinitesimal times, but is not sufficient to prove positivity
on timescales of order the memory time or longer.
In \cite{Wonderen00} there is a general argument for the preservation of 
positivity, however we believe that this argument is only correct at 
infinite times.
Ref.~\cite{Wonderen00} also has an interesting discussion of 
conditions for positivity of a two-level system modelled 
(phenomenologically) by a Bloch equation with time dependent parameters. 
However in our experience, 
master equations with negative coupling constants
do not take the form of Bloch equations
(see the model in Section~\ref{sect:2level}).
\bibitem{footnote:factorized-state}
Some authors claim that factorized states are only relevant to those 
extremely rare situations in which the system-environment coupling is 
switched-on at $t=0$ (i.e. no coupling for all $t<0$).  
This is incorrect. 
Even with finite coupling for 
all $t$ (including $t<0$), a projective measurement on the system
at $t=0$
prepares a given system state (with the experimenter knowing with certainty
what this state is). This forces the total density matrix
to be $\rhosys(t=0) \otimes \hat \rho_{\rm env}$, where $\rhosys(t=0)$
is the given (pure) system state.
\bibitem{footnote:relation-S-to-A}
This relationship between $S(\om)$ and $A(\om)$ can be shown as follows.
At thermal equilibrium 
$\hat{\rho}_{\rm env} = \sum_n P_n |n \rangle \langle n|$ where
$|n\rangle$ is an eigenstate of $\hat{\cal H}_{\rm env}$, and
$P_n \propto \exp[ -E_n/k_{\rm B}T]$.  In this case
$S(\om)$ and $A(\om)$ are $\propto \sum_{nm} P_n |x_{mn}|^2 
[\delta (E_m-E_n-\om)\pm  \delta (E_m-E_n+\om)]$,
where $+(-)$ is for $S(\om)$ ($A(\om)$).
Using $\sum_n = \int \rmd E' \nu(E')$
and then writing energy as the average and difference of $E_m$ and $E_n$,
we find $S(\om)$ and $A(\om)$ are 
$\propto \e^{\om/2k_{\rm B}T} \pm \e^{-\om/2k_{\rm B}T}$.
\bibitem{footnote:maxime}
The rotating-wave approximation fails if the 
dephasing rate is faster than $(E_{i'}-E_{j'}-E_i+E_{j'})$
with $i'\neq i$ and $j'\neq j$.
Thus it always fails
if there are 
(i) degeneracies (i.e.~$E_{i'}=E_i$), 
or (ii) two gaps of equal magnitude  (i.e.~$E_{i'}-E_{j'}=E_i-E_j$)
\cite{Maxime-paper}.
\bibitem{footnote:three-levels}
Consider a 3-level system whose
density matrix is diagonal with the three following
elements; $\rho_{11}=(1+x)/3$, $\rho_{22}=(1-x)/3$, $\rho_{33}=1/3$.
For $1<x \leq \sqrt{3}$, this density-matrix is unphysical
because it is not positive (the probability $\rho_{22}<0$),
yet the purity $P \leq 1$.
\bibitem{footnote:polar}
This is easily shown by writing $s_x,s_y,s_z$ in polar coordinates,
and showing that the maxima of the resulting function are at
$(\theta,\phi)=(\pi/2, \half Bt)$ and $(\theta,\phi)=(\pi/2, \pi +\half Bt)$.
\end{thebibliography}
\end{document}